\documentclass[12pt,onecolumn]{IEEEtran}

\usepackage{times}

\usepackage{graphicx}
\usepackage[english]{babel}
\usepackage[latin1]{inputenc}
\usepackage{amssymb}
\usepackage{url}

\usepackage[ruled]{algorithm2e}

\usepackage{setspace}

\graphicspath{{Plots/}}

\newcommand{\ie}{{\em i.e.}}

\newcommand{\cc}{\ensuremath{\mbox{cc}}}
\newcommand{\rc}{\ensuremath{\mbox{rc}}}

\newcommand{\actor}{{\em actors-movies}}
\newcommand{\authoring}{{\em authoring}}
\newcommand{\occurrence}{{\em occurrences}}
\newcommand{\ptp}{{\em peer-to-peer}}

\newcommand{\plots}[3][0.34]{\begin{figure}[!h]
\hspace*{1cm}
\actor
\hspace*{2cm}
\authoring
\hspace*{2.8cm}
\occurrence
\hspace*{2cm}
\ptp\\
\centering
\includegraphics[scale=#1]{actor/top/#2.eps} 
\includegraphics[scale=#1]{coauthoring/top/#2.eps} 
\includegraphics[scale=#1]{cooccurrence/top/#2.eps} 
\includegraphics[scale=#1]{p2p/top/#2.eps}\\
\includegraphics[scale=#1]{actor/bot/#2.eps} 
\includegraphics[scale=#1]{coauthoring/bot/#2.eps} 
\includegraphics[scale=#1]{cooccurrence/bot/#2.eps} 
\includegraphics[scale=#1]{p2p/bot/#2.eps}
\caption{#3
First row: for top nodes.
Second row: for bottom nodes.
}
\label{fig_#2}
\end{figure}
}

\newcommand{\noteperso}[1]{
\vskip 0.5cm
\centerline{\fbox{\begin{minipage}{15cm} #1\end{minipage}}}
\vskip 0.5cm}

\begin{document}

\begin{center}

{\LARGE Basic Notions for the Analysis of Large
\vskip 0.2cm
Affiliation Networks / Bipartite Graphs}

\vskip 0.3cm

Matthieu Latapy,\footnote{
 LIAFA -- CNRS and Universit\'e Paris~7 --
 2, place Jussieu, 75005 Paris, France --
 {\tt latapy@liafa.jussieu.fr}}
Cl\'emence Magnien\,\footnote{
 CREA -- CNRS and Ecole Polytechnique --
 5, rue Descartes, 75005 Paris, France --
 {\tt magnien@shs.polytechnique.fr}}
and
Nathalie Del Vecchio\,\footnote{
 LARGEPA -- Université Paris~2 --
 13, avenue Bosquet, 75007 Paris, France --
 {\tt nathdelvecchio@yahoo.com}
}
\vskip 1cm

\noteperso{\centering \small preprint -- comments are welcome -- {\tt latapy@liafa.jussieu.fr} -- \today}
\vskip 1cm

\end{center}



\begin{center}

 \textbf{Abstract}

 \vskip 0.3cm

 \begin{minipage}{15cm}

 Many real-world complex networks actually have a bipartite nature:
 their nodes may be separated into two classes, the links being
 between nodes of different classes only. Despite this, and despite the
 fact that many ad-hoc tools have been designed for the study of special
 cases, very few exist
 to analyse (describe, extract relevant information) such networks in a
 systematic way. We propose here an extension of the most basic
 notions used nowadays to analyse classical complex networks to the
 bipartite case. To achieve this, we introduce
 a set of simple statistics, which we discuss by comparing their
 values on a representative set of real-world networks and on their
 random versions.

 \end{minipage}

\end{center}

\vskip 1cm

\section*{Introduction.}

A bipartite graph is a triplet $G=(\top,\bot,E)$ where $\top$ is the
set of {\em top} nodes, $\bot$ is the set of {\em bottom} nodes, and
$E \subseteq \top \times \bot$ is the set of links. The difference
with {\em classical} graphs lies in the fact that the nodes are in
two disjoint sets, and that the links always are between a node of one
set and a node of the other. In other words, there cannot be any link
between two nodes in the same set.

Many real-world complex networks may be modeled  naturally by a
bipartite graph. Let us cite for example the actors-movies network,
where each actor is linked to the movies he/she played in
\cite{watts98collective,newman01random},
authoring networks, where the authors are linked to the paper they
signed \cite{newman01scientific1,newman01scientific2},
occurrence networks, where the words occurring in a book
are linked to the sentences of the book they appear in \cite{ferrer01small},
company board networks, where the board members are linked to the
companies they lead \cite{robins04small,conyon04small,battistion03statistical},
and peer-to-peer exchange networks in which peers are linked to the data
they provide/search \cite{lefessant04clustering,voulgaris04exploiting,iptps05,iwdc04}. These networks
are often called {\em affiliation networks}, or {\em two-mode networks}.

Although there is nowadays a significant amount of notions
and tools to analyse classical networks, there is still a lack of such
results fitting the needs for analysing affiliation networks.
In such cases, one generally has to transform the affiliation network
into a classical one and/or to introduce ad-hoc notions. In the first
case, there is an important loss of information, as well as other problems that
we detail below (Section~\ref{sec_proj}). In the second case, there is
a lack of rigor and generality, which makes the relevance of the obtained
results difficult to evaluate.

The aim of this contribution is to provide a set of simple statistics
which will make it possible and easy to analyse real-world affiliation
networks (or at least make the first step towards this goal) while keeping their bipartite nature.

To achieve this, we will first present an overview of the basic notions
and methodologies used in the analysis of classical (as opposed to affiliation)
networks. We will then show how people usually transform bipartite networks
into classical networks in order to be able to analyse them with the
tools designed for this case. This will lead us to a description of the
state of the art, then of the methodology used in this paper. Finally, we
will present and evaluate the statistics we propose for the analysis of
affiliation networks.

Before entering in the core of this contribution, let us notice that
we only deal here with simple\,\footnote{This means that we do not allow
loops (links from a node to itself) nor multiple links between two given
nodes. This is classical in complex networks studies:
loops are managed separately, if some occur, and multiple links are
generally encoded as link weighs, or simply ignored.}, undirected,
unweighted, static networks.
Considering directed, weighted, and/or dynamic networks is
out of the scope of this paper; we will discuss this further in the
conclusion. Moreover, in all the cases we will consider here (and
in most real-world cases), the graph has a giant connected component,
\ie\ there exists a path in the graph from almost any node to any
other. In the following, we will make our statistics on the whole
graph everywhere this makes sense, but we will restrict ourselves to
the largest connected component where this is necessary (namely for distance
computations). Again, this is classical in the literature and has no
significant impact on our results.

\section{Classical notions.}
\label{sec_notions}

Let us consider a (classical) graph $G = (V,E)$, where $V$ is the set of
nodes and $E \subseteq V \times V$ is the set of links. We will denote
by $N(v) = \{u\in V,\ (u,v) \in E\}$ the {\em neighbourhood} of a node $v$,  the
elements of $N(v)$ being the {\em  neighbours} of $v$. The number of nodes
in $N(v)$ is the {\em degree} of $v$: $d^o(v) = |N(v)|$.

The most basic statistics describing such a graph are its
size $n = |V|$, its number of links $m = |E|$, and its average
degree $k = \frac{2m}{n}$.
Its density $\delta(G) = \frac{2m}{n(n-1)}$, \ie\ the number of
existing links divided by the number of possible links, also
is an important notion. It is nothing but the
probability that two randomly chosen (distinct) nodes are linked together.

Going further, one may define the distance between two nodes in the
graph as the minimal number of links one has to follow to go from
one node to the other. Note that this only make sense if there is a
path between the two nodes, \ie\ if they are in the same connected
component. As explained above, in all the paper, we will only consider distances between
the nodes in the largest connected component (and we will give its
size). Then, the average distance of the graph, $d(G)$,
is nothing but the average of the distances for all pairs of nodes
in the largest connected component.

The statistics described above are the ones we will call the {\em basic}
statistics. The next one is not so classical. It is the degree
distribution, \ie\ for all  integer $i$ the fraction $p_i$ of nodes of
degree $i$. In other words, it is the probability that a randomly chosen
node has degree $i$. One may also observe the correlations between
degrees, defined as the average degree of the neighbours of nodes of
degree $i$, for each integer $i$. Other notions concerning degrees
have been studied, like assortativity \cite{newman03mixing} for
instance, but we do not detail this here.

The last kind of statistics we will discuss here aims at capturing a notion
of overlap: it measures the probability that two nodes are linked together,
provided they have a neighbour in common. In other words, it is the probability
that any two neighbours of any node are linked together. This may be done using
two slightly different notions, both called {\em clustering coefficient},
among which there often is a confusion in the literature\,\footnote{Some
authors make a difference by calling the first notion {\em clustering coefficient}
and the second one {\em transitivity ratio}, but we prefer to follow the
most classical conventions of complex network studies here.}.
Both will be useful in the following therefore we discuss them precisely here.

The first one computes the probability, for any given node chosen at random,
that two neighbours of this node are linked together. It therefore relies
on the notion of clustering coefficient for any node $v$ of degree at least
$2$, defined by:
$$\cc_{\bullet}(v) = \frac{|E_{N(v)}|}{\frac{|N(v)|(|N(v)|-1)}{2}}
                  = \frac{2|E_{N(v)}|}{d^o(v)(d^o(v)-1)}$$
where $E_{N(v)} = E \cap (N(v)\times N(v))$ is the set of links between
neighbours of $v$.
In other words, $\cc_{\bullet}(v)$ is the probability that two
neighbours of $v$ are linked
together. Notice that it is nothing but the density of the neighbourhood
of $v$, and in this sense it captures the local density.
The clustering coefficient of the graph itself is the average of
this value for all the nodes:
$$\cc_{\bullet}(G) = \frac{\sum_{v\in V} \cc_{\bullet}(v)}{|\{v\in V,\ d^o(v)\ge2\}|}.$$

\medskip

\noindent
One may define directly another notion of clustering coefficient of $G$ as a whole as
follows:
$$\cc_{\vee}(G) = \frac{3 N_{\Delta}}{N_{\vee}}$$
where $N_{\Delta}$ denotes the number of triangles, \ie\ sets of three
nodes with three links in $G$, and $N_{\vee}$ denotes the number  of
connected triples, \ie\ sets of three nodes with at least two links,
in $G$. This notion of clustering is slightly different from the previous
one since it gives the probability, when one chooses two links with one
extremity in common, that the two other extremities are linked together.

Both notions have their own drawbacks and advantages. The first one has the
advantage of giving a value for each node,  which makes it possible to
observe the  distribution of this value and the correlations between this
value and the degree, for instance. It however has the drawback of
reducing the role of high degree nodes. Moreover, importantly, these
definitions capture slightly  different notions, which may both be relevant
depending on the context. We will therefore use both notions
in the following. This is why we introduced two different notations,
namely $\cc_{\bullet}$ and $\cc_{\vee}$, which emphasises the fact that
one is centered on nodes and the other is centered on pairs of links
with one extremity in common.

One may consider many other statistics to describe complex networks. Let
us cite for instance centrality measures, various decompositions, and
notions capturing the ability of each node to spread information
in the network. See
\cite{wasserman94social,albert02statistical,newman03structure,bornholdt03book,Brandes:Network_Analysis}
for surveys from different perspectives. We will not consider here such
statistics. Instead, we will focus on the most simple ones, described above,
because they play a central role in recent complex network studies as we will
explain in the next section.

\section{Classical complex networks.}
\label{sec_classical}

Many complex networks have been
studied in the literature, ranging from technological networks (power
grids, internet) to social ones (collaboration networks, economical
relations),  or from biological ones (protein interactions, brain
topology) to linguistic ones (cooccurrence networks, synonymy networks).
See \cite{wasserman94social,albert02statistical,newman03structure,bornholdt03book,Brandes:Network_Analysis} and references therein for detailed examples.
It appeared recently \cite{watts98collective,albert02statistical,newman03structure,bornholdt03book}
that most of these large real-world
complex networks have several nontrivial properties in common.
This section is devoted to an overview and a discussion of these
properties (based on the definitions given in previous section),
on which the rest of the paper will rely. We will use the same
notations as in Section~\ref{sec_notions}.

We are concerned here with large complex networks, which
means that $n$ is large. In most real-world cases, it appeared that
$m$ is of the same order of magnitude as $n$, \ie\ the average degree $k$
is  small compared to $n$. Therefore, the density generally is very
small: $\delta(G) = \frac{kn}{n(n-1)} \sim \frac{k}{n}$,  which
is close to $0$ since $n$ is much larger than $k$ in general.
We will always suppose we are in this case in the following.

It is now a well known fact that the average distance in real-world
complex networks is in general very small ({\em small-world} effect),
even in very large ones, see for instance \cite{milgram67small,watts98collective}.
This is actually true in most graphs, since a small amount of randomness
is sufficient to ensure this, see for instance \cite{watts98collective,kleinberg00navigation,kleinberg99smallworld,bollobas01random,erdos59random}.
This property, though it may have important consequences and should be
taken into account, should therefore not be considered as a significant
property of a given network (see Section~\ref{sec_methodo}).

Another issue which received recently much attention, see for instance
\cite{faloutsos99powerlaw,barabasi99emergence}, is the fact that the degree distribution of most real-world
complex networks is  highly heterogeneous, often well fitted by a power
law: $p_k \sim k^{-\alpha}$ for an exponent $\alpha$ generally between
$2$ and $3.5$. This means  that, despite most nodes have a low
degree, there exists nodes with a very high degree. This implies in
general that the average degree is not a significant property, bringing
much less information than the exponent $\alpha$ which is a measurement
of the heterogeneity of degrees.

If one samples a random network with the same size (\ie\ as many
nodes and links)
as a given real-world one\,\footnote{We consider here a network chosen
uniformly at random among the ones having this size,
using typically the Erd\"os and R\'eny\'\i{} model \cite{bollobas01random,erdos59random}.}, thus with
the same density, then the obtained degree
distribution is qualitatively different: it follows a Poisson law.
This means that the heterogeneous degree distribution is not a trivial
property, in the sense that it makes real-world complex networks very
different from most graphs (of which a random graph is typical).
The degree correlations and other properties on degrees, however,
behave differently depending on the complex network under concern.

Going further, the clustering coefficients (according to both definitions)
are quite large in most real-world complex networks: despite most pairs
of nodes are not linked together (the density is very low), if two nodes
have a neighbour in common then they are linked together with a probability
significantly higher than $0$ (the local density is high). However, the
clustering coefficient distributions, their correlations with degrees,
and other properties related to clustering, behave differently depending
on the complex network under concern.

If, as above, one samples a random graph with the same size as an
original one then the
two definitions of clustering coefficients are equivalent and equal to
the density. The clustering coefficients therefore are very low in this
case. If one samples a random graph
with the same number of nodes {\em and} the very same degree
distribution\,\footnote{We consider here a network chosen
uniformly at random among the ones having this number of nodes and this
degree distribution, using typically the {\em configuration} model
\cite{bender78asymptotic,bollobas01random,molloy95critical,molloy98size,viger05efficient}.}
then the clustering coefficients still are very small, close to $0$ \cite{newman03structure}.
Clustering coefficients therefore capture a property of networks which
is not a trivial consequence of their degree distribution.

Finally, it was observed that the vast majority of large real-world complex
networks have a very low density, a small average distance, a highly
heterogeneous degree distribution and high clustering coefficients. These
two last properties make them very different from random graphs (both
purely random and random with prescribed degree distribution). More
subtle properties may be studied, but until now no other one appeared to
be a general feature of most real-world complex network. The properties
described here therefore serve as a basis for the analysis of real-world
complex networks, and so we will focus on them in the following. Our aim will be
to define and discuss their equivalent for affiliation networks~/~bipartite
graphs.

\section{Projection.}
\label{sec_proj}

\begin{figure}[h!]
\centering
\includegraphics[scale=0.8]{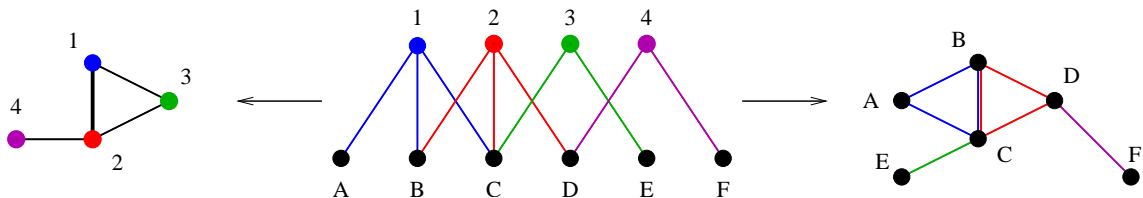}
\caption{An example of bipartite graph (center), together with
its $\top$-projection (left) and its $\bot$-projection
(right).}
\label{fig_proj}
\end{figure}

Let us now consider a large affiliation network modeled as a bipartite
graph $G=(\top,\bot,E)$. The \mbox{$\bot$-projection} of $G$ is the graph
$G_{\bot}=(\bot,E_{\bot})$ in which two nodes (of $\bot$) are linked
together if they have at least one neighbour in common (in $\top$) in $G$:
$E_{\bot} = \{ (u,v),\ \exists\, x \in \top:\ (u,x)\in E\mbox{ and } (v,x)\in E\}$.
The $\top$-projection $G_{\top}$ is defined dually.
See Figure~\ref{fig_proj} for an example.

In order to be able to use the many notions defined on classical networks,
and to compare a particular network to others, one generally transforms
an affiliation network into its $\bot$-projection, often called the
one-mode version of the network. This was typically done for the affiliation
networks we presented in the introduction: the actors-movies network is
transformed into its $\bot$-projection where two actors are linked together
if they acted together in a movie \cite{watts98collective}; the authoring networks are
transformed into their $\bot$-projections, \ie\ coauthoring networks where
two authors are linked together if they signed a paper together \cite{newman01scientific1,newman01scientific2,newman01random};
the occurrence networks are transformed into their $\bot$-projections,
\ie\ cooccurrence networks where two words are linked if they appear in
the same sentence \cite{ferrer01small}; the company board networks are
transformed into their $\bot$-projections where two persons are linked
together if they are member of a same board \cite{robins04small,conyon04small,battistion03statistical}; and the peer-to-peer
exchange networks are transformed into their $\bot$-projections where
two data are linked together if they are provided/searched by a same
peer \cite{lefessant04clustering,voulgaris04exploiting,iptps05,iwdc04}.

This approach is of course relevant since the projections under study
make sense, and also encode relevant information. Moreover, this
allows the study of affiliation networks using the powerful tools
and notions provided for classical networks.
We however argue that in most cases  there would
be a significant gain in considering the bipartite version of the data.
The main reasons are as follows.
\begin{itemize}
\item
Most importantly, there is much information in the bipartite structure
which may disappear after projection. For instance, the fact that two
actors played in many movies together, and the size of these movies,
brings much information which in not available in the projection, in
which they are simply
linked together. This loss  of information is particularly clear when
one notices that there are many bipartite graphs which lead to the same
projection (while each bipartite graph has only one $\top$- and one
$\bot$-projection), see \cite{ipl04,caan04}.
The fact that much important information is encoded in
the bipartite structure is a central point which we will illustrate
all along this paper.
\item
Notice that each top node of degree $d$ induces $\frac{d(d-1)}{2}$ links
in the $\bot$-projection, and conversely. This induces an inflation of the
number of links when one goes from a bipartite gaph to its projection, see
Table~\ref{tab_size}. In our examples, this is particularly true for
peer-to-peer: the number of links reaches more than $10$ billions
in the $\bot$-projection, which needs more than $80$ GigaBytes of
central memory to be stored using classical (compact) encodings
(while the original affiliation network needs less than $500$ MegaBytes).
This is a typical case in which the huge number of links induced by
the projection is responsible for
limitations on the computations we are able to handle on the graph in practice.
\begin{table}[h!]
\centering
\begin{tabular}{l|r|r|r|r}
                    & actors-movies & authoring & occurrences & peer-to-peer\\
\hline
Number of links in $G$ &
1\mbox{,}470\mbox{,}418 &
45\mbox{,}904 &
183\mbox{,}363 &
55\mbox{,}829\mbox{,}392 \\
Number of links in $G_{\bot}$ &
15\mbox{,}038\mbox{,}083 &
29\mbox{,}552 &
392\mbox{,}066 &
10\mbox{,}142\mbox{,}780\mbox{,}673
 \\
Number of links in $G_{\top}$ &
20\mbox{,}490\mbox{,}112 &
134\mbox{,}492 &
51\mbox{,}405\mbox{,}275 &
1\mbox{,}085\mbox{,}217\mbox{,}140 \\
\end{tabular}
\caption{Number of links in affiliation networks and their
projections, for the four examples we will describe in
Section~\ref{sec_methodo}.}
\label{tab_size}
\end{table}
\item
Finally, some properties of the projection may be due to the projection
process rather than the underlying data itself. For instance, it is
shown in \cite{newman01random,ipl04,caan04} that when considering
the projection of a random
bipartite graph, one observes high clustering coefficients. Therefore,
high clustering coefficients in projections may not be viewed as significant
properties: they are consequences of the bipartite nature of the
underlying affiliation network. Likewise, the projection may lead to very
dense networks, even if the bipartite version is not dense; this is
particularly the case here for the $\top$-projection of occurrences.
\end{itemize}

\noindent
One way to avoid some of these problems is to use a {\em weighted} projection.
For instance, the weight of a link $(u,v)$ between two bottom nodes in the
weighted $\bot$-projection may be defined as the number of (top) neighbours $u$
and $v$ have in common in the bipartite graph. Other definitions may be
considered: each top node may contribute to each link it induces in the
$\bot$-projection in a way that decreases with its degree, for instance.
In all cases, and despite such an approach is relevant and promising,
one still looses a significant amount of information, and one transforms
the problem of analysing a bipartite structure into the problem of analysing
a weighted one, which is not easier. Indeed, despite the fact that important
progress has recently be done in this direction
\cite{barrat04architecture,barthelemy05characterization,newman04analysis},
much remains to be done to be able to analyse precisely the structure of
weighted networks.

Our aim is this paper is to provide an alternative to the projection
approach, leading to a better understanding of affiliation networks.
It must however be clear that (weighted) projection approachs also lead to
significant insight, and we consider that the two approaches should
be used as complementary means to understand in details the properties
of affiliation networks.

\section{State of the art.}
\label{context}

Affiliation networks have been studied in an amazingly wide variety
of context. Let us cite for instance
company boards~\cite{robins04small,conyon04small,battistion03statistical,newman01random},
sport teams~\cite{bonacich72technique,onody04complex},
movie actors~\cite{watts98collective,newman01random},
human sexual relations~\cite{ergun02human,lind05cycles},
attendance to political events~\cite{faust02scaling},
financial networks~\cite{caldarelli03emergence,dahui05bipartite,garlaschelli04scale,young98structural},
recommandation networks~\cite{perugini03connection},
theatre performances~\cite{agneessens04choices,uzzi04collaboration},
politic ativism~\cite{boudourides04networks},
student course registrations~\cite{holme04korean},
word cooccurrences~\cite{dhillon01coclustering,veronis95large},
file sharing~\cite{iamnitchi04small,lefessant04clustering,voulgaris04exploiting,iptps05,iwdc04},
and scientific authoring~\cite{roth05epistemic,morris05construction,newman01scientific1,newman01scientific2,newman04who}.

These studies are made in disciplines as various as social sciences,
computer science, linguistics and physics, which makes the literature
very rich. In all these contexts, scientists
face affiliation networks which they try to analyse, with various motivations
and tools. They all have one feature in common: they insist on the fact that
the bipartite nature of their data plays an important role, and should
be taken into account. They also emphasise the lack of notions and tools
for doing so.

Because of this lack of relevant notions and tools, most authors have no
choice but to consider the most relevant projection of their affiliation
network. This leads for instance to studies of interlocks between companies,
see for instance~\cite{robins04small,conyon04small}, studies of
coauthoring networks, see for instance~\cite{newman01scientific1,newman01scientific2,newman04who},
or studies of exchanges between peers in peer-to-peer systems,
see for instance~\cite{lefessant04clustering,voulgaris04exploiting,iptps05,iwdc04}.

Many authors realise that this approach is not sufficient, and try to use
the bipartite nature of their data. This is generally done by combining
the use of projections and the use of basic bipartite statistics, mostly
degrees. For instance, one studies the coauthoring relations (typically
a projection) and the distributions of the number of papers signed by
authors and of the number of authors of papers (\ie\ the bipartite
degree distributions, see Section~\ref{sec_basic})~\cite{newman04who}.
Authors may also consider weighted projections, see for
instance~\cite{battistion03statistical,morris05construction,iwdc04,iptps05,iamnitchi04small,newman04who},
which has advantages and drawbacks, as discussed above,
Section~\ref{sec_proj}.

Going further, some authors introduce bipartite notions designed for the
case under study. This is often implicit and restricted to very basic
properties, like the case of degree distributions cited above (which
essentially capture the size of {\em events} and the number of events
in which {\em persons} or {\em objects} are involved,
in most cases). But some authors introduce more subtle notions,
like notions of overlap~\cite{bonacich72technique},
clustering~\cite{borgatti97network,robins04small,lind05cycles},
centrality measures~\cite{faust97centrality},
degree correlations~\cite{peltomaki05correlations},
and others~\cite{young98structural,ergun02human,caldarelli03emergence,perugini03connection,iamnitchi04small,borgatti97network,robins04small,lind05cycles}.
Most of these notions are ad hoc and specific to the case under
study, but some of them actually are very general or may be
generalised. One of our central aims here is to give a complete
and unified framework for the most general of these notions. We will cite
appropriate references when the notions we will discuss have already
been considered previously.

As already said, a different and interesting approach is developed
in~\cite{newman01random,ipl04,caan04}. The authors study the expected
properties of the projections given the properties (namely
the degree distributions) of the underlying bipartite graph.
They show in particular that the expected clustering coefficient
in the projections is large, and give an efficient estimation
formula; this means that a high clustering coefficient in a
projection may be seen as a consequence of the underlying
bipartite structure rather than a specific property of the
network. Conversly, if the clustering coefficient of the
projection is different from the expected one, it means that
the underlying bipartite structure has nontrivial properties
responsible for it. These properties should therefore
be further analysed. Our aim here is to propose notions
and tools for such an analysis.
This approach has been used with profit in several cases, for
instance~\cite{newman01random,newman02random,conyon04small,uzzi04collaboration}.

\medskip

Finally, a significant effort has already been made to achieve
the goal we have
here, or similar goals: some studies propose general approaches for the
analysis of affiliation networks. This is for instance the case
of~\cite{faust97centrality}, focused on centrality measures,
of~\cite{breiger74duality}, which proposes to consider both projections
and compare them, and of~\cite{bonacich72technique}, which studies in depth
the notion of overlap.

Let us cite in particular~\cite{borgatti97network}, which has the
very same aim as we have here, but makes quite different choices.
For instance, the authors deal with visualisation, whereas we do
not here. On the other hand, they consider rather small networks,
while we are particularly interested in large ones. They use
a matrix approach, which is very powerful but not suited for large
cases. This also leads them to consider some properties, like
(betweenness) centrality for instance, which can hardly be computed
on very large networks. Finally, they do
not use the comparison with random graphs, central to our
contribution (see  Section~\ref{sec_methodo}),
which probably reflects the fact that this method was not as usual
in 1997 as it is now. It is interesting to see that, although the
initially claimed aim is very similar, the final contributions
are very different.

Other researchers propose formalisms suited for the analysis
of affiliation networks, often based on a generalisation of well
known models.
Let us cite Galois lattices~\cite{roth05epistemic},
correspondence analysis~\cite{roberts00correspondence,faust06using},
extensions of blockmodels~\cite{borgatti92regular,doreian04generalized}
and p* models~\cite{skvoretz00logit,faust02scaling,agneessens04choices}
and a particularly original approach based on boolean
algebra~\cite{bonacich78using}.

Therefore, there already exists quite an impressive amount of work
on affiliation networks, and on methods for their analysis.
However, we observe that many of the approaches proposed previously,
though very relevant, are
hardly applicable to {\em large} networks, typically networks with
several hundreds of thousands  nodes.
Moreover, they often rely on quite complex
notions and formalisms, which are difficult to handle for people
only interested in analysing a given network.
Finally, none of them consists
in a generalisation of the notions nowadays widely used to analyse
classical (as opposed to affilliation) networks, outlined in
Section~\ref{sec_notions}.

We propose here such a contribution. We design simple notions
and methods to analyse very large affiliation networks, which
could be used as a first step in particular studies.
These methods may then be extended to fit the
details of particular cases, and we explain how to do so.
Moreover, they are not
only extensions of classical notions; we go further by proposing
new notions designed specifically for the bipartite case.
Our approach may also be applied to smaller networks, as long as
they are not too small (typically thousands of nodes).

As explained  above, the topic has a deep  interdisciplinary
nature. In order to make our techniques usable by a wide audience,
we give a didactic
presentation and we focus on basic notions. Let us insist however
on the fact that this presentation is rigorous and formal, and, as
will appear all along the paper, the results are sufficient to
bring a significant amount of information on a given network.

Finally, we
insist on the fact that analysing properly and in details a given
network is a difficult task, which may be handled using different
methods. There is no unique way to obtain relevant information
and results in such cases. Moreover, much resides in the
interpretations made from the outputs of these approaches.
All the ones we have cited above, and the one we propose
here, should therefore be seen as complementary rather than concurrent.

Let us conclude this section by noticing that, because of the wide
dispersion of contributions due to the interdisciplinary nature of
the topic (and the fact that it received continuous attention since
several decades), we certainly missed some references. We however expect
that the ones we have cited span well the contributions on the topic.

\section{Methodology and data.}
\label{sec_methodo}

The methodology we follow to develop tools for the analysis of complex
networks of various kinds, mainly used since the end of the
90's, is the one sketched in Section~\ref{sec_classical}. It relies on the
introduction of statistical parameters aimed at capturing a given
feature of networks under concern, an then on the comparison of the
behaviours of real-world networks concerning these parameters as
compared to random ones. The underlying principle is that a  parameter
which behaves  similarly on real-world and random networks is just
a property of {\em most} networks (of which random networks are
representatives) and so, though it may play an important role, it
should not be considered as surprising and meaningful concerning
the description of the real-world network. Instead, one generally
looks for properties which make real-world networks different from
most networks.

Our contribution here relies on this methodology. Namely, we will
define statistical parameters aimed at capturing properties of
bipartite graphs, and then evaluate the relevance of these parameters
by comparing their values on random bipartite graphs and on real-world
affiliation networks.

Just like one considers purely random graphs and random graphs
with prescribed degree distributions in the case of classical
networks, we will use both purely random bipartite graphs and
random bipartite graphs with prescribed degree distributions. Such
graphs are constructed easily by extending the classical case,
see for instance \cite{newman01random,ipl04}. We provide a program
generating such graphs at \cite{urlgenerators}.
Note that these models (both the classical and bipartite versions) generate 
graphs that are not necessarily {\em simple}: they may contain some
loops and multiple links. There are however very few such links, and
simply removing them generally has no impact on the results. This
is what is generally done in the literature, and we will follow this
convention here: in our context, it cannot have a significant
impact\,\footnote{One may also use the methods described in
\cite{viger05efficient} to obtain directly simple (connected) graphs,
but this is more intricate, and unnecessary in our context.}.

Notice also that the properties of random graphs may be formally evaluated,
see for instance \cite{newman01random,caan04}. We will however focus on practical aspects
here and leave these evaluations for further work, see Section~\ref{sec_conclu}.

In order to complete our comparison between random and real-world
cases, we also need a set of real-world affiliation networks. We chose the
following four instances, which correspond to the examples given in the
introduction and have the advantage of spanning well the
variety of cases met in practice:
\begin{itemize}
\item
the {\em actors-movies} network as obtained from the {\em Internet Movie
Data Base} \cite{imdburl} in 2005, concerning
$n_{\bot}=127\mbox{,}823$ actors and
$n_{\top}=383\mbox{,}640$ movies, with
$m=1\mbox{,}470\mbox{,}418$ links;
\item an {\em authoring} network obtained from the online {\em arXiv}
preprint repository \cite{arxivurl}, with
$n_{\top}=19\mbox{,}885$ papers,
$n_{\bot}=16\mbox{,}400$ authors, and
$m=45\mbox{,}904$ links;
\item an {\em occurrence} graph obtained from a version of the Bible \cite{bibleurl}
which contains
$n_{\bot}=9\mbox{,}264$ words and
$n_{\top}=13\mbox{,}587$ sentences with
$m=183\mbox{,}363$ links;
\item a {\em peer-to-peer} exchange network obtained by registering all
the exchanges processed by a large server during 48 hours \cite{iptps05,iwdc04},
leading to
$n_{\top}=1\mbox{,}986\mbox{,}588$ peers,
$n_{\bot}=5\mbox{,}380\mbox{,}546$ data, and
$m=55\mbox{,}829\mbox{,}392$ links;
\end{itemize}
We provide these data, together with the programs computing the
statistics described in this paper, at \cite{dataurl}. The key point here
is that this dataset spans quite well the variety of context in which
large affiliation networks appear, as well as the variety of data
sizes.

Let us insist on the fact that our aim here is not to derive conclusions
on these particular networks: we only use them as real-world instances to
illustrate the use of our results and to discuss their generality. This is
why we do not detail more the way they are gathered and their
relevance to any study. This is discussed in the corresponding references
and is out of the scope of this paper.

\section{Basic bipartite statistics.}
\label{sec_basic}

The basic statistics on bipartite graphs are direct extensions of the
ones on classical graphs. One just has to be careful with the fact that
some classical properties give birth to twin bipartite properties while
others must be redefined.

Let us consider a bipartite graph $G=(\top,\bot,E)$. We denote by
$n_{\top} = |\top|$ and $n_{\bot} = |\bot|$ the numbers of top and
bottom nodes, respectively. We denote by $m = |E|$ the number of links
in the network. This  leads to a top average degree
$k_{\top} = \frac{m}{n_{\top}}$ and a bottom one
$k_{\bot} = \frac{m}{n_{\bot}}$. One may obtain the average degree in
the graph $G'=(\top\cup\bot,E)$ as
$k = \frac{2 m}{n_{\top}+n_{\bot}}
   = \frac{n_{\top} k_{\top} +  n_{\bot} k_{\bot}}{n_{\top}+n_{\bot}}$.
Finally, we  obtain the bipartite density
$\delta(G) = \frac{m}{n_{\top} n_{\bot}}$, \ie\ the fraction of existing
links with repect to possible ones. Note that this is different from the density
of $G'$:
$\delta(G') = \frac{2m}{(n_{\top}+n_{\bot})(n_{\top}+n_{\bot}-1)}$,
which is much lower.

Concerning the average distance (again, we restrict distance
computations\,\footnote{Distance computations are expensive;
the exact value cannot be computed in a reasonable amount of time for data
of the size we consider here. Instead, we approximate the average by computing
the average distance from a subset of the nodes to all the others, this
subset being large enough to ensure that increasing it does not improve our
estimation anymore, which is a classical method. All other computations are exact.}
to the largest connected component, which contains the vast majority of nodes,
see Table~\ref{tab_basic}), there is no crucial difference except
that one may be interested by the average distance between top nodes
and between bottom nodes, $d_{\top}$ and $d_{\bot}$. These values  may
be significantly different but one may expect that they are very close
since a path between two top (resp. bottom) nodes is nothing but a path
between bottom (resp. top) nodes with two additionnal links. Notice
that there is no simple way to derive the average distance $d$ in
$G'$ from the bipartite statistics $d_{\bot}$ and $d_{\top}$.

\begin{table}[h!]
\centering
\begin{tabular}{l|cc|cc|cc|cc}
  & \multicolumn{2}{c|}{actors-movies}
  & \multicolumn{2}{c|}{authoring}
  & \multicolumn{2}{c|}{occurrences}
  & \multicolumn{2}{c}{peer-to-peer}\\
  & real & random
  & real & random
  & real & random
  & real & random\\
\hline

$n_{\top}$ &
$127\mbox{,}823$& \em idem &
$19\mbox{,}885$& \em idem &
$13\mbox{,}587$& \em idem &
$1\mbox{,}986\mbox{,}588$ & \em idem \\

$n_{\bot}$ &
$383\mbox{,}640$& \em idem &
$16\mbox{,}400$& \em idem &
$9\mbox{,}264$& \em idem &
$5\mbox{,}380\mbox{,}546$ & \em idem \\

$m$ &
$1\mbox{,}470\mbox{,}418$& \em idem &
$45\mbox{,}904$& \em idem &
$183\mbox{,}363$& \em idem &
$55\mbox{,}829\mbox{,}392$ & \em idem \\

$k_{\top}$ &
$11.5$& \em idem &
$2.3$& \em idem &
$13.5$& \em idem &
$28.1$ & \em idem \\

$k_{\bot}$ &
$3.8$& \em idem &
$2.8$& \em idem &
$19.8$& \em idem &
$10.4$ & \em idem \\

$k$ &
$5.7$& \em idem &
$2.5$& \em idem &
$16.0$& \em idem &
$15.2$ & \em idem \\

$\delta$ &
$0.000030$& \em idem &
$0.00014$& \em idem &
$0.0015$& \em idem &
$0.0000052$ & \em idem \\

$\top$ giant &
$124\mbox{,}414$& $125\mbox{,}944$&
$16\mbox{,}209$& $18\mbox{,}512$&
$13\mbox{,}579$& $13\mbox{,}587$&
$1\mbox{,}986\mbox{,}343$ & $1\mbox{,}426\mbox{,}978$\\

$\bot$ giant &
$374\mbox{,}511$& $381\mbox{,}431$&
$11\mbox{,}654$& $14\mbox{,}607$&
$9\mbox{,}246$& $9\mbox{,}264$&
$5\mbox{,}380\mbox{,}507$ & $5\mbox{,}054\mbox{,}689$\\

$d_{\top}$ &
$6.8$& $5.3$&
$13.1$& $9.3$&
$3.1$& $3.0$&
$5.3$ & $5.0$\\

$d_{\bot}$ &
$7.3$& $5.8$&
$13.9$& $9.9$&
$3.8$& $3.7$&
$5.4$ & $4.9$\\

$d$ &
$7.2$& $5.8$&
$13.5$& $9.6$&
$3.4$& $3.2$&
$5.3$ & $4.9$

\end{tabular}
\caption{Basic bipartite statistics on our four examples and on random
bipartite graphs with the same size.}
\label{tab_basic}
\end{table}

The values obtained for each of these basic properties on our four examples,
together with values obained for random bipartite networks with the same
size, are given in Table~\ref{tab_basic}.
It appears clearly that our examples may be
considered as large networks with small average degrees, compared to their
size. The density therefore is small. Moreover, the average distance is
also small. These basic properties are very similar to what is  observed
on classical networks: both classical and affiliation large real-world
complex networks are sparse and have a small average distance, and in
both contexts this is also true on random graphs.

\section{Bipartite statistics on degrees.}
\label{sec_bip_degrees}

The notion of degree distribution has an immediate extension to the
bipartite case. We denote by $\bot_i$ the fraction of nodes in $\bot$
having degree $i$ and by $\top_i$ the fraction of nodes in $\top$
having degree $i$, and then call $(\bot_i)_{i\ge0}$ the bottom degree
distribution and $(\top_i)_{i\ge0}$ the top one.

\plots{deg_distrib}{
Degree distributions in our four real-world affiliation networks.
}

The top and bottom degree distributions of our four examples are given
in Figure~\ref{fig_deg_distrib}. One may observe on these plots that
the bottom degree distributions are very heterogeneous and well fitted by
power laws (of various exponents). This is true in particular for the
occurrences graph, which is a well known fact for a long time
\cite{zipf32selective}: the frequency of occurrences of words in a
text generally follows a particular kind of power law, named {\em Zipf}
law.
Instead, the shape of the top degree
distribution depends on the case under concern: whereas it is well fitted
by a power law in the peer-to-peer and actors-movies cases, it is far from a power
law in the authoring and occurrences cases.
This is due to the fact that papers have a limited number of
authors (none has one hundred authors for instance), and likewise
sentences have a limited number of words. Moreover, the number of very short
sentences also is not huge. In these two cases, one can hardly conclude 
that the top degrees are very heterogeneous.

We  finally conclude that,
even if heterogeneity is present on at least one side of an affiliation
network, this is not generally true for both sides. This separates real-world
affiliation networks into two distinct classes, which should be taken
into account in practice. This also confirms that considering the bipartite
statistics brings significant information as compared to the projections,
which exhibit power law degree distributions in all cases.

\medskip

Let us now compare these real-world statistics with random graphs.
If one generates purely random bipartite graphs of the same size as
the ones considered here, the ($\top$ and $\bot$)
degree distributions are Poisson laws. Therefore, the heterogenity of
some  degree distributions is not present, and even in the cases where
the distributions are not very heterogeneous they do not fit the random
case. We will therefore compare in the following our real-world affiliation
networks to random bipartite graphs with the same size and the same (top
and bottom) degree distributions.

The next natural step is to observe possible correlations between top
and bottom degrees. In order to do this, we plot in Figure~\ref{fig_deg_cor}
the average degree of neighbours of nodes as a function of their degree,
both for top and bottom nodes, separately. In other words, for each integer
$i$ we plot the average degree of all nodes which are neighbours of a node
of degree $i$. We plot the same values obtained for random graphs of the same size
and same degree distributions.

\plots{deg_cor}{
Degree correlations in our four real-world affiliation networks,
and in random bipartite graphs of the same size and
same degree distributions.
}

In the cases of actors-movies and peer-to-peer, the plots for the random cases are
close to horizontal lines, showing that there are no correlations between
a node degree and the average degree of its neighbours: this last value is
independent of the node degree. In both cases, however, the real-world
network displays nontrivial correlations. In the case of actors-movies, for instance,
the average degree of neighbours of bottom nodes (the lower-left corner plot
in Figure~\ref{fig_deg_cor}) decreases with the node degree. In other words,
if an actor plays in many movies then he/she tends to play in smaller
movies (in terms of the number of involved actors). Such nontrivial
observations may be made on the other plots for actors-movies and peer-to-peer as
well.

In the cases of authoring and occurrences, the plots for the random
graphs are nontrivial: they grow for the top statistics, and are far
from smooth for the bottom ones. Here again, the real-world cases exhibit
significantly different behaviours, at least for the top statistics, thus
demonstrating that these behaviours are nontrivial and related to intrinsinc
properties of the underlying networks. Detailing this however is out of
the scope of this paper. The key point here is to have evidence of the
relevance of these statistics.

\medskip

Notice that, despite they already bring much information, the statistics
observed until now are almost immediate extensions of the classical ones.
One may wonder if the bipartite nature of the networks under concern may
lead to entirely new notions concerning degrees.
We propose one below, with its variants.

Let us consider a
node $v$ in a bipartite graph $G=(\top,\bot,E)$, and let us denote
by $N(N(v))$ the nodes at distance $2$ from $v$, not including $v$,
called {\em distance $2$ neighbours} of $v$. We will suppose that
$v$ is a top node, the other case being dual. Notice that 
$N(N(v)) \subseteq \top$, and actually $N(N(v))$ is nothing but 
$N(v)$ in the $\top$-projection $G_{\top}$. The integer $|N(N(v))|$ therefore
plays a central role in the projection approach, since it is the degree of
$v$ in $G_{\top}$.

But there are several ways for $v$ to be linked to the nodes in
$N(N(v))$, this information being lost during the projection. The two
extreme cases occur when $v$ is linked to only one node $u$ in $\bot$,
with $N(u) = N(N(v))$, or when $v$ is linked to $|N(N(v))|$ nodes in
$\bot$, each being linked to only one other node in $\top$. Of course,
intermediate cases may occur, and the actual situation may be observed
by plotting the correlations between the degree of nodes $v$, \ie\
$|N(v)|$, and their
number of distance $2$ neighbours, $|N(N(v))|$.
These statistics therefore offer a way to study how node degrees in
the projection
appear, and to distinguish between different behaviours. For instance, they
make it possible to say if a given author has many coauthors because
he/she writes many papers or if he/she writes papers with many authors.
Such an information is not available in the projection of the authoring
affiliation network.

\plots{deg_2_cor}{
Correlations of the number of distance $2$ neighbours
with node degrees in our four examples, and in random bipartite
graphs with the same size and degree distributions.
}

The plots in Figure~\ref{fig_deg_2_cor}
show that, as one may have guessed, the number of distance $2$ neighbours
of a node grows with its degree; more precisely, it generally grows as
a power of the degree (the plots follow straight lines in log-log
scale), and actually almost linearly. This is in conformance with the
intuition that the number of distance $2$ neighbours should be close to
the degree of the node times the average degree of its neighbours. In
the random cases, this leads to very straight plots (except in the top
plot of occurrences). The real-world plots are quite close to the random
ones, with a few notable exceptions: the slope of the plot is significantly
different for the top plot of peer-to-peer, the real-world plots often
are significantly below the random ones for large degrees, and they are
in general slightly lower than the random ones even for small degrees.
This means that there is some redundancy in the neighbourhoods: whereas
in random cases the number of distance $2$ neighbours is close to the sum of
the degrees of the direct neighbours, in real-world cases the direct
neighbours have many neighbours in common and so the number of distance $2$
neighbours is significantly lower. This is an important feature of
real-world complex networks, that we will deepen in the next sections.

\section{Bipartite clustering and overlap.}
\label{sec_bip_clustering}

Whereas there were quite direct extensions of the basic statistics and
the ones on degrees to the bipartite case, the notion of clustering
coefficient does not make any sense in itself in this context. Indeed,
it relies on the enumeration of the triangles in the graphs, and there
can be no triangle in a bipartite graph. We will therefore have to
discuss the features captured by the classical clustering coefficients
in order to propose bipartite extensions.

\begin{figure}[h!]
\centering
\includegraphics[scale=0.7]{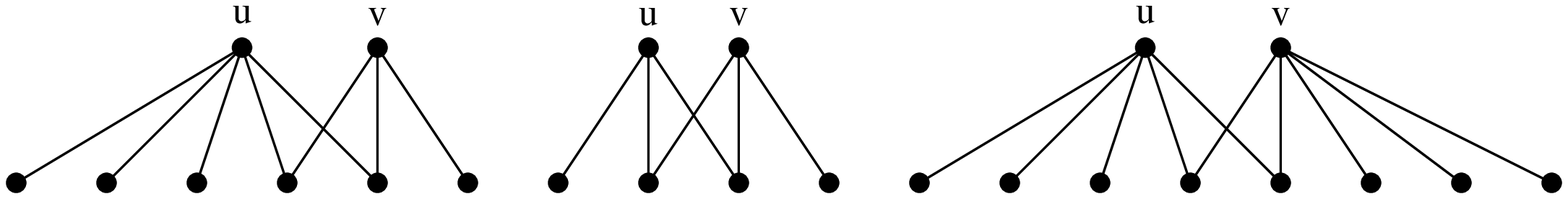}
\caption{Examples of bipartite clustering coefficients, and interpretations.
Left: a case in which $\cc_{\bullet}(u,v) = \frac{2}{6} = 0.333\cdots$ is quite
small, despite the fact that $u$ and $v$ have two neighbours in common, due to
the fact that the union of their neighbours is quite large; on
the countrary, $\cc_{\underline\bullet}(u,v) = \frac{2}{3} = 0.666\cdots$ is
quite large, revealing that one of the neighbourhoods is almost included in
the other; the value of $\cc_{\overline\bullet}(u,v) = \frac{2}{5} = 0.4$
indicates that this may be due to the fact that one of the nodes has a
high degree. The situation is different
in the case at the center: all clustering coefficients are quite high
(resp. $0.5$, $0.666\cdots$, and $0.666\cdots$), indicating
that there is not only an important overlap, but that this overlap concerns
a significant part of each neighbourhoods (and thus the two nodes have similar
degrees). On the right, the two nodes have a small clustering coefficient
$\cc_{\bullet}(u,v) = \frac{2}{8} = 0.25$, and the fact that the
value of $\cc_{\underline\bullet}(u,v) = \frac{2}{5} = 0.4$ remains quite
small indicates that this is not due to the fact that one of the two nodes
has a very high degree compared to the other one. If ones considers larger
degree nodes, then the difference between {\em small} and {\em high} values
is clearer, but the figure would be unreadable.}
\label{fig_clustering}
\end{figure}

Both definitions of classical clustering coefficients capture the fact
that when two nodes have something in common (one neighbour) then they
are linked together with a probability much higher than two randomly
chosen nodes. Conversly, they capture the fact that when two nodes
are linked together then they probably have neighbours in common. In
other words, they capture correlations between neighbourhoods.
We will use this point of view here and define a first notion of clustering
coefficient defined for pairs of nodes (in the same set $\top$ or $\bot$):
$$ \cc_{\bullet}(u,v) = \frac{|N(u) \cap N(v)|}{|N(u) \cup N(v)|}.$$
This is the most direct generalisation of the classical notion, and it
was already suggested
in \cite{borgatti97network}, and explicitely used in \cite{iptps05}
in the context of peer-to-peer exchange analysis.
It captures the overlap between neighbourhoods of nodes: if
$u$ and $v$ have no neighbour in common then $\cc_{\bullet}(u,v)=0$. If they have the
same neighbourhood, then $\cc_{\bullet}(u,v)=1$. And if their neighbourhoods partially
overlap then the value is in between, closer to $1$ when the overlap is
large compared to their degrees.
See Figure~\ref{fig_clustering} for an illustration.

This definition however has several drawbacks. The first one is the fact
that it defines a value for {\em pairs} of nodes. One may want to capture
the tendency of {\em one} particular node to have its neighbourhood included
in the ones of other nodes. To achieve this, one may simply define the
clustering coefficient of one node as the average of its clustering
coefficients with other nodes. We however do not include in this averaging
the pairs for which the overlap is empty\,\footnote{
As a consequence, the obtained value
will never be $0$, but it may be very small. Notice also that the clustering
coefficient is not defined for nodes $v$ such that $N(N(v)) = \emptyset$ (recall
that, by definition, $v\not\in N(N(v))$.}:
most nodes have disjoint
neighbourhood, which does not bring information. Like in the classical case,
we want to measure the implication of the fact of having one neighbour in
common on the rest of the neighbourhoods. We finally obtain:
$$\cc_{\bullet}(u) = \frac{\sum_{v\in N(N(u))}\cc_{\bullet}(u,v)}{|N(N(u))|}$$
One may then observe the distribution of these values, their correlations
with degrees, etc. One may also define the clustering coefficient of the
top (resp. bottom) nodes, denoted by $\cc_{\bullet}(\top)$ (resp.
$\cc_{\bullet}(\bot)$) as the average of this value over top (resp.
bottom) nodes. The average over the all graph, denoted by
$\cc_{\bullet}(G)$, can then be obtained
easily: $\cc_{\bullet}(G) = \frac{n_{\top}\cc_{\bullet}(\top)+n_{\bot}\cc_{\bullet}(\bot)}{n_{\top}+n_{\bot}}$.
We will discuss the obtained values below, see Table~\ref{tab_bip_cc}.

\medskip

The notion of clustering coefficient discussed until now is an extension
of the first classical one. It captures the fact that a node which has a
neighbour in common with another node generally has a significant portion
of neighbours in common with it. There is another way to capture this, similar
to the second definition of classical clustering coefficient, is to measure
the probability that, given four nodes with three links, they actually
are connected with four links (all the possible bipartite ones):
$$ \cc_{\sf N}(G) = \frac{2N_{\Join}}{N_{\sf N}}$$
where $N_{\Join}$ is the number of quadruplets of nodes with four links in $G$,
and $N_{\sf N}$ is the number of quadruplets of nodes with at least three.
This extension of the second notion of classical clustering coefficient
was already proposed in \cite{robins04small} in the context of company
board networks.
We will discuss the obtained values below, see Table~\ref{tab_bip_cc}.

\medskip

The two notions above generalise the classical definitions of clustering
coefficients. Capturing the overlap between neighbours may however need
more precision. Suppose that degrees are heterogeneous in the network,
as it is often the case (Section~\ref{sec_bip_degrees}), and consider two nodes
$u$ and $v$. If one of these nodes has a high degree and the other has not,
then $\cc_{\bullet}(u,v)$ will necessarily be small. This will be true
even if one of the neighbourhoods is entirely included in the other. One
may however want to capture this, which can be done using the following
definition:
$$ \cc_{\underline\bullet}(u,v) = \frac{|N(u) \cap N(v)|}{\min\left(|N(u)|,|N(v)|\right)}.$$
One may define dually:
$$ \cc_{\overline\bullet}(u,v) = \frac{|N(u) \cap N(v)|}{\max\left(|N(u)|,|N(v)|\right)}.$$
See Figure~\ref{fig_clustering} for an illustration.
These two notions, called min- and max-clustering, were introduced first
in \cite{iptps05}. The first one emphasises  on the fact that small neighbourhoods
may intersect significantly large ones; it is equal to $1$ whenever one of the
neighbourhoods is included in the other. The second one emphasises on the fact
that neighbourhoods (both small or large ones) may overlap very significantly:
it is $1$ only when the two neighbourhoods are the same and it tends to
decreases rapidly if the degree of one of the involved nodes increases.
It captures the fact that nodes with {\em similar} degrees
have high neighbourhood overlaps.

With these definitions, one may define
$\cc_{\underline\bullet}(v)$, 
$\cc_{\underline\bullet}(\top)$, 
$\cc_{\underline\bullet}(\bot)$, 
$\cc_{\underline\bullet}(G)$, 
$\cc_{\overline\bullet}(v)$,
$\cc_{\overline\bullet}(\top)$,
$\cc_{\overline\bullet}(\bot)$,
and
$\cc_{\overline\bullet}(G)$
in a way similar to the one used above for
$\cc_{\bullet}(v)$,
$\cc_{\bullet}(\top)$,
$\cc_{\bullet}(\bot)$,
and
$\cc_{\bullet}(G)$.
The distributions and various correlations may then be observed.

\begin{table}[h!]
\centering
\begin{tabular}{l|cc|cc|cc|cc}
  & \multicolumn{2}{c|}{actors-movies}
  & \multicolumn{2}{c|}{authoring}
  & \multicolumn{2}{c|}{occurrences}
  & \multicolumn{2}{c}{peer-to-peer}\\
  & real & random
  & real & random
  & real & random
  & real & random\\
\hline

$\cc_{\bullet}(\top)$ &
$0.064$& $0.046$&
$0.29$& $0.27$&
$0.066$& $0.066$&
$0.056$ & $0.019$\\

$\cc_{\bullet}(\bot)$ &
$0.36$& $0.20$&
$0.31$& $0.25$&
$0.065$& $0.038$&
$0.076$ & $0.074$\\

\hline

$\cc_{\sf N}(G)$ &
$0.0082$& $0.00024$&
$0.079$& $0.00012$&
$0.053$& $0.048$&
$0.0094$ & $0.00019$\\
\hline

$\cc_{\underline\bullet}(\top)$ &
$0.24$& $0.23$&
$0.56$& $0.56$&
$0.19$& $0.20$&
$0.27$ & $0.24$\\

$\cc_{\underline\bullet}(\bot)$ &
$0.81$& $0.79$&
$0.73$& $0.70$&
$0.64$& $0.61$&
$0.39$ & $0.42$\\

\hline

$\cc_{\overline\bullet}(\top)$ &
$0.087$& $0.062$&
$0.36$& $0.34$&
$0.097$& $0.097$&
$0.074$ & $0.024$\\

$\cc_{\overline\bullet}(\bot)$ &
$0.37$& $0.21$&
$0.33$& $0.26$&
$0.069$& $0.041$&
$0.091$ & $0.089$\\


\end{tabular}
\caption{Bipartite clustering statistics on our four examples and on random
bipartite graphs with the same size and same degree distributions.}
\label{tab_bip_cc}
\end{table}

We give in Table~\ref{tab_bip_cc} the values obtained for our four examples
together with the values obtained for random bipartite graphs with
same size and degree distributions (the values for purely
random bipartite graphs are similar). It appears clearly  that
the notions we introduced capture different kinds of overlaps between
neighbourhoods. However, except for $\cc_{\sc N}(G)$, the obtained values
are not very different on random graphs and on real-world networks.
This indicates that these statistics do not capture a very significant
feature of real-world complex networks, which will discuss this further
below. Instead, the obtained values for $\cc_{\sc N}(G)$ is significantly
larger on real-world networks than on random graphs, which shows that it
captures more relevant information.

\plots{cc}{
Cumulative distributions of the various clustering coefficients
in our four real-world affiliation networks.
}

\medskip

We show in Figure~\ref{fig_cc} the cumulative distributions of $\cc_{\bullet}(v)$,
$\cc_{\underline\bullet}(v)$ and $\cc_{\overline\bullet}(v)$ for our four
examples, \ie\ for each value $x$ on the horizontal axis the ratio of all
the nodes having a value lower than $x$ for these statistics.
Before entering in the discussion of these plots, notice that, by
definition, we have
$
\cc_{\bullet}(v)
\le
\cc_{\overline\bullet}(v)
\le
\cc_{\underline\bullet}(v)
$ for any $v$. Therefore, the lower plots in
each case of Figure~\ref{fig_cc} is the one of $\cc_{\underline\bullet}(v)$,
the upper is the one for $\cc_{\bullet}(v)$ and the one for
$\cc_{\overline\bullet}(v)$ is in between.

More interesting, the plots exhibit quite different behaviours. In several
cases (in particular top of actors-movies, occurrences and peer-to-peer, as well
as bottom of occurrences and peer-to-peer) the plots for
$\cc_{\overline\bullet}(v)$ and $\cc_{\bullet}(v)$ grow very rapidly and
are close to $1$ almost immediately. This means that the values of these
statistics are very small, almost $0$, for most nodes: in these cases,
the neighbours of nodes have a small intersection, compared to the union
of their neighbourhoods. However, in several cases, the plots for
$\cc_{\underline\bullet}(v)$ grow much less quickly, and remain lower than
$1$ for a long time. In several cases, it is even significantly lower
than $1$ by the end of the plot, meaning that for an important number of nodes
the value of $\cc_{\underline\bullet}(v)$ is equal to $1$: almost
$10\%$ in the case of top of actors-movies, almost $20\%$ in the cases of top
authoring and bottom of peer-to-peer, and more thant $40\%$ in the
case of bottom of occurrences. This means that, despite overlaps are
in general small compared to their possible value, the neighbourhoods of
many low-degree nodes significantly or even completely overlap with
other nodes neighbours.

Other cases display a very different behaviour: in both top and bottom
plots of authoring, and in bottom of actors-movies, it appears clearly that a
significant number of nodes have a large value for
$\cc_{\underline\bullet}(v)$, $\cc_{\bullet}(v)$ and
$\cc_{\overline\bullet}(v)$. This means that node neighbours overlap
significantly, and that this is not only a consequence of the fact
that low degree nodes have their neighbourhoods included in the ones
of other nodes.

\medskip

\plots{ccr}{
Cumulative distributions of the $\cc_{\bullet}$ clustering coefficient
in our four real-world affiliation networks, and in
random bipartite graphs of the same size and
same degree distributions.
}

Again, our aim here is not to discuss in detail the specificities of
each case, but to give evidence of the fact that these statistics
have nontrivial behaviours and capture significant information. It is
clear from the discussion above that the three notions of clustering
captured by $\cc_{\underline\bullet}(v)$, $\cc_{\bullet}(v)$ and
$\cc_{\overline\bullet}(v)$ are different, and give complementary
insight on the underlying network properties. One may however be
surprised by the fact that $\cc_{\bullet}(v)$ often is very small,
which we deepen now by comparing its behaviours on real-world cases
and on random ones, see Figure~\ref{fig_ccr}\,\footnote{
For clarity and to avoid long discussions on specific behaviours,
which is out of our scope here, we only compare the real-world and the
random behaviours of $\cc_{\bullet}(v)$ (not of the two other notions of
clustering coefficients).}.

In these plots, it appears clearly that, except in the case of
bottom of actors-movies, the plots of the real-world values and of
the random ones are quite similar. This means that, concerning the
values of $\cc_{\bullet}(v)$, real-world graphs are not drastically
different from random ones (they however have slightly higher
values of $\cc_{\bullet}(v)$ in most cases). In other words, this
statistics does not capture very significant information, according to
the methodology described in Section~\ref{sec_methodo}. This is due
to the fact that the low degree nodes (which are numerous in our
networks) have with high probability their neighbours in common
with high degree nodes; by definition, this induces a low value
for $\cc_{\bullet}(v)$, and even lower for $\cc_{\overline\bullet}(v)$.
This is true by construction for random graphs, and the plots above
show that this is mostly true for real-world networks also, which
was not obvious.

Similar
conclusions follow from the study of $\cc_{\overline\bullet}(v)$,
but the study of $\cc_{\underline\bullet}(v)$ leads to the
opposite conclusion: an important
number of nodes have their neighbourhood included in the one of other
(large degree) nodes, as already discussed, which happens much more
rarely in random graphs. We do not detail these results here, since
they do not fit in the scope of this paper. Instead, we will
propose a new statistics in the next section that has several
advantages on the clustering coefficients discussed here and does
not have their drawbacks.

\medskip

Before turning to this other statistics, let us
observe the correlations between node degrees and their clustering
coefficient. Again, for clarity and to maintain the paper within a reasonable
length, we focus on $\cc_{\bullet}(v)$ and its comparison with the random
case. See Figure~\ref{fig_deg_cc_dot_cor}.

\plots{deg_cc_dot_cor}{
Correlations of the $\cc_{\bullet}(v)$ clustering coefficient
with node degrees in our four examples, and in random bipartite
graphs with the same size and degree distributions.
}

The values for the random graphs are below the ones for the real-world cases
(or they coincide at some points), in all plots. This shows that the
value of $\cc_{\bullet}(v)$ are larger in real-world cases than
in random ones, but the difference is small, which confirms the
observations above. More interestingly, it appears clearly that in
most cases $\cc_{\bullet}(v)$ decreases as a power of the degree of $v$
(straight line in log-log scale). In other words, the clustering
coefficient of low degree nodes is quite large, but the one of large
degree nodes is very small, like in random graphs.

\section{The notion of redundancy.}
\label{sec_bip_redundancy}

In the previous section, we discussed several ways to extend the classical
notions of clustering coefficient to the bipartite case.
One may wonder if the bipartite nature of the networks under concern may
lead to new, specific notions, just like we observed
concerning degrees in Section~\ref{sec_bip_degrees}.
Moreover, one may want to capture
the notion of overlap concerning {\em one} particular node; in previous
section, this was only possible by averaging the value obtained for
a possibly large number of pairs of nodes.
This section answers this: it is devoted to a new notion aimed at capturing
overlap in bipartite networks, in a node-centered fashion.

\begin{figure}[h!]
\centering
\includegraphics[scale=0.7]{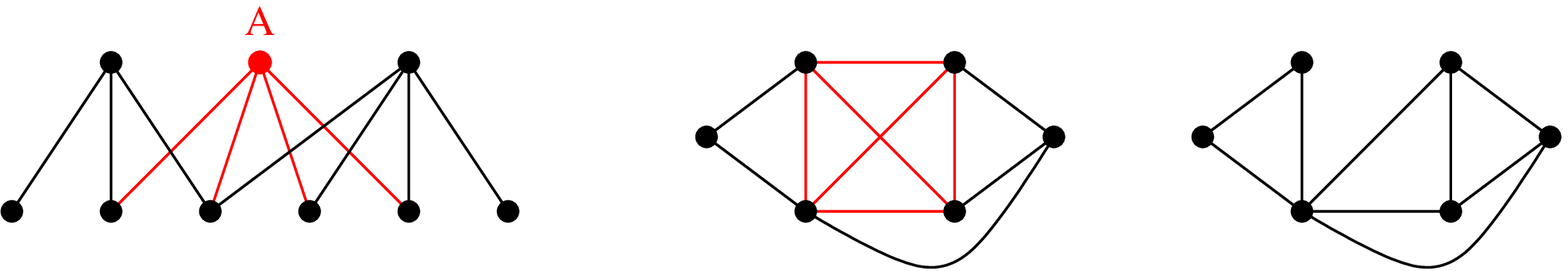}
\caption{Example of redundancy computation.
From left to right: a bipartite graph, its $\bot$-projection, and
the $\bot$-projection obtained if the node $A$ is first removed. Only two
links disappear, leading to $\rc(A) = \frac{4}{6} = 0.666\cdots$.}
\label{fig_redundancy}
\end{figure}

First notice that neighbourhood overlaps correspond to links
which are obtained in several ways during the projection, and that these
links cannot be distinguished one from another in the projection. They also
reveal the fact that, among all the links induced by a node of a
bipartite graph in the projection, many (and possibly all) may actually
be induced by others too. In  other words, if we remove this node in the
bipartite graph then the projection may be only slightly changed (or even
not at all). This can be captured by the following parameter,  which we
call the {\em redundancy coefficient} of $v$:
$$\rc(v) = \frac{|\{ \{u,w\}\subseteq N(v),\ \exists\, v'\not= v,\ (v',u)\in E \mbox{ and } (v',w)\in E\}|}{\frac{|N(v)|(|N(v)|-1)}{2}}.$$
In other words, the redundancy coefficient of $v$ is the fraction of
pairs of neighbours of $v$ linked to another node than $v$. In the projection,
these nodes would be linked together even if $v$ were not there, see
Figure~\ref{fig_redundancy}; this
is why we call this the redundancy. If it is equal to $1$ then
the projection would be exactly the same without $v$; if it is $0$ it
means that none of its neighbours would be linked together in the
projection.

Again, we can derive from this definition the ones of $\rc(\top)$,
$\rc(\bot)$ and $\rc(G)$, as well as distributions and correlations.
We give in Table~\ref{tab_bip_rc} the values obtained for our four
examples and for comparable random graphs. It appears clearly from
these values that, except in the case of occurrences, the redundancy
coefficient is much larger in real-world networks than in random
graphs, and that it actually is very large: in peer-to-peer, for
instance, on average half the pairs of peers that have a common
interest for a
given data also have a common interest for another data. These values
are much larger than the ones for the clustering coefficients in
the previous section, see Table~\ref{tab_bip_cc}, and the difference they
make between random graphs and real-world networks is much more
significant. To this regard, it may be considered as a better
generalisation of clustering coefficients in classical networks
(as opposed to affiliation ones) than the bipartite clustering
coefficients defined in Section~\ref{sec_bip_clustering}.

The case of occurrences is different: the projections on both sides
are very dense, which is  very particular as already noticed. The
redundancy coefficient therefore is huge, but this is not because of a
property of how the neighbourhoods overlap: this is a direct consequence
of the high density of the projections. In such a case, the redundancy
coefficient is meaningless, and we will therefore not discuss this
case any further in this section; simply notice that the redundancy
coefficient has similar behaviours in such graphs and in their random
equivalent.

\begin{table}[h!]
\centering
\begin{tabular}{l|cc|cc|cc|cc}
  & \multicolumn{2}{c|}{actors-movies}
  & \multicolumn{2}{c|}{authoring}
  & \multicolumn{2}{c|}{occurrences}
  & \multicolumn{2}{c}{peer-to-peer}\\
  & real & random
  & real & random
  & real & random
  & real & random\\
\hline
$\rc(\top)$ &
$0.26$& $0.014$&
$0.38$& $0.0016$&
$0.80$& $0.74$&
$0.31$ & $0.011$\\

$\rc(\bot)$ &
$0.25$& $0.011$&
$0.33$& $0.00037$&
$0.83$& $0.75$&
$0.50$ & $0.069$\\

\end{tabular}
\caption{The redundancy coefficient for our four examples and for random
bipartite graphs with the same size and same degree distributions.}
\label{tab_bip_rc}
\end{table}

We show in Figure~\ref{fig_rc_distrib} the distributions of $\rc(v)$
for our four examples together with plots for comparable random graphs.
These plots confirm that the redundancy coefficient captures
a property that makes real-world complex networks different from random
ones: in all the cases except occurrences, the value of this coefficient
in random graphs is almost $0$ for all nodes (both top and bottom);
instead, in real-world networks it is significantly larger, and equal to
$1$ for a large portion of the nodes. This last fact is not surprising
since $\cc_{\underline\bullet}(v)=1$ implies $\rc(v)=1$ for all nodes $v$.

\plots{rc_distrib}{
Cumulative distributions of the redundancy coefficient
in our four real-world affiliation networks,
and in random bipartite graphs of the same size and
same degree distributions.
}

However, the redundancy coefficient has a much wider range of values than
$\cc_{\underline\bullet}(v)$, which generally is close
to $0$ or $1$, see Figure~\ref{fig_cc}. Moreover, the
redundancy coefficient captures a different property: in the case of
actors-movies, for
instance, it does not only mean that a significant number of movies have
a cast that is a sub-cast of another movie (as captured by
$\cc_{\underline\bullet}(v)$), but that when two actors
act together in a movie then there often exists (at least) another movie in which
they also act together. Both are interesting, and complementary, but the
redundancy coefficient certainly captures a more general feature.

\medskip

Let us now observe the correlations between node redundancy coefficient
and their degree, plotted in Figure~\ref{fig_deg_rc_cor}. In these
plots, except for occurrences, the plots for the random graphs
coincide with the x-axis, which confirms that the values of node
redundancy in random graphs are very small, independently of node
degrees. Real-world cases, on the countrary, exhibit nontrivial
behaviours. In most cases, the redundancy decreases with the degree,
which is not surprising since the number of links needed in the
projection so that the redundancy of a node is large grows with the
square of its degree. However, the redundancy remains large even
for quite large degrees: it is close to $0.15$ for nodes of
degree $30$ for top nodes in actors-movies, for instance, meaning that
among the $435$ possible pairs of neighbours of these nodes, on
average $65$ are linked to another top node in common. This has
a very low probability in random graphs. Likewise, one may notice
that some very high degree nodes have a very large redundancy
coefficient in several cases, which also is a significant information.

\plots{deg_rc_cor}{
Correlations of redundancy coefficient with node degrees
in our four real-world affiliation networks,
and in random bipartite graphs of the same size and
same degree distributions.
}

One may push further the study of the redundancy, for instance by
counting how many nodes have an overlap with a given
one, and so may be responsible for its high redundancy. This
is nothing but the degree of the node in the appropriate
projection, which emphasises once again that our approach may be
combined with the one based on projection with benefit, as argued in
Section~\ref{sec_proj}.

\section{Conclusion and perspectives.}
\label{sec_conclu}

The core contribution of this paper is a  set of rigorous and coherent
statistical properties usable as a basis for the analysis  of large
real-world affiliation networks. These statistics go from the very basics
(size, distances, etc) to subtle ones (typically various clustering
coefficients and their correlations with degrees). Let us insist on the
fact that we do not only extend classical  notions to the bipartite case,
but also  develop new notions which make sense only in this context. 
Moreover, the proposed approach avoids projection of affiliation networks
into classical ones, which makes it possible to grab much richer information.
We hope that this unified framework and discussion will help significantly
people involved in analysis of such networks.

A first conclusion driven from the computation of these statistics on
four representative real-world examples is that, just like real-world
classical complex  networks, they have nontrivial properties in common
which make them very different from random networks. In particular,
there is a high heterogeneity between degrees of nodes of at least one
kind, and there are significant overlaps between neighbourhoods.
Concerning this last property, we show that immediate extensions of
the classical notions of clustering coefficients are not sufficient
to make the difference between real-world networks and random graphs;
we propose the notion of {\em redundancy} as a relevant alternative.
Overall, these
facts are strikingly close to what is met in classical networks and should
play a  similar role. Conversly, we observed many properties which behave
differently depending on the affiliation network under concern, which
may be used to describe a particular instance in more details.

Notice that these contributions do not only concern the affiliation
networks themselves, but also their projection: keeping the bipartite
nature of the data makes it possible to obtain more precise information
on the projection itself. For instance, statistics on degrees make it
possible to separate high degree nodes in the projection into two
distinct classes: the ones which are linked to many nodes in the affiliation
network, and the ones linked to nodes of high degree in the affiliation
network. This kind  of analysis could be deepened using clustering
and redundancy notions.

Going further, one may use the notions we introduced here to define new
relevant statistics on classical networks. Indeed, any (classical) graph
$G=(V,E)$ may be seen as a bipartite graph $G'=(V,V,E)$ where the links
are between two {\em copies} of $V$. The statistics we studied here may
then be computed on this bipartite graph, leading to new insight on the
original graph $G$.

\medskip

There are many directions to improve and continue the work presented here.
Among them, let us cite the analytic study of the parameters we propose,
which can in particular be done  using the techniques in \cite{newman01arbitrary}.
One might prove in this way the expected behaviour of these parameters and
deepen their understanding. Another direction is the developement of
models of affiliation networks capturing the properties met in practice.
Just like it is the case for classical networks, much can be done
concerning degrees, but very little is known concerning the modeling of
clustering and redundancy. Finally, applying these results to practical
cases and giving
precise interpretations of their meanings in these different contexts
would probably help in designing other relevant notions.

Let us conclude by noticing that large complex network analysis is only at
its  beginning, though much has been done in recent years on classical
networks. However, most real-world networks are directed, weighted, 
labelled, hybrid, and/or evolve during time.
Some work has recently been done concerning weighted networks \cite{barrat04architecture,barthelemy05characterization,newman04analysis}, and
we propose here a contribution concerning affiliation networks. However,
there is still much to do to be able to analyse efficiently these various
kinds of networks. Extending the notions we discussed here to the case
of multipartite graphs (nodes are in several disjoint sets, with links
between nodes in different sets only) would be a step further in this
direction.

{

\small

\medskip
\subsubsection*{Acknowledgments}
We warmly thank
Arnaud Bringé,
Dominique Cardon,
Pascal Cristofoli,
Nicolas Gast,
Jean-Loup Guillaume,
Christophe Prieur,
Camille Roth
and
Fabienne Venant,
as well as the {\em SocNet} community,
for their precious comments and help during the preparation of this contribution.
This work was funded in part by the
PERSI ({\em Programme d'\'Etude des R\'eseaux Sociaux de l'Internet}) project
and by the AGRI ({\em Analyse des Grands R\'eseaux d'Interactions}) project.

}

\bibliographystyle{plain}
\bibliography{Biblio/xbib,Biblio/xbib_bip,Biblio/xbib_new}

\end{document}